\documentstyle[preprint,aps,epsf]{revtex}
\begin{document}
\preprint{APCTP/98-16}
\draft
\title{Topological Ward Identity and\\
Anti-de Sitter Space/CFT Correspondence
}
\author{Taejin Lee\cite{tlee}}
\address{Asia Pacific Center for Theoretical Physics, 207-43
Cheongryangri-dong\\
Dongdaemun-gu, Seoul 130-012, Korea \\ and \\
Department of Physics, Kangwon National University, 
                       Chuncheon 200-701, Korea} 
\date{\today}
\maketitle
\begin{abstract}
The dual relationship between the supergravity in the anti-de 
Sitter(AdS) space and the superconformal field theory is discussed 
in the simplest form We show that a topological Ward identity holds 
in the three dimensional Chern-Simons gravity. 
In this simple case the proposed dual relationship 
can be understood as the topological Ward identity. 
Extensions to the supersymmetric theories and higher dimensional ones
are also briefly discussed. 
\end{abstract}
\pacs{PACS number(s): 04.60.K, 11.10.Kk, 11.25.Hf}
%


The duality has been one of the most powerful tools to understand the 
strong coupling regimes of various quantum systems . The dual 
relationship recently conjectured by Maldacena \cite{mal97} brings us to explore 
a new avenue in this direction. It connects the supergravity in the 
anti-de Sitter (AdS) space and the superconformal field theory on the boundary.
The proposed duality has been refined in \cite{gkp98,witten981}
and many interesting facets of the duality have been discussed in refs.
\cite{itzh98,gunaydin981,horow98,ferrara,rey98,mal98,witten982,freed98,vish,vafa,suss98,gross}.
In particular the AdS/CFT correspondence has been useful to address the various
subtle issues associated with the large $N$ limit of super-Yang-Mills theory
in terms of the classical geometry of the supergravity in the AdS space 
\cite{rey98,mal98,witten982,freed98,vafa,suss98,gross}.
Although the AdS/CFT correspondence has been discussed in the various settings,
it certainly needs to be further refined to be a reliable practical
tool to study the large $N$ super-Yang-Mills theory and the nonperturbative 
D-brane dynamics. 
In this paper we will show that the AdS/CFT 
correspondence holds as a consequence of the Ward identity in some simple 
examples. It clearly exhibits the topological nature of the duality.

We will begin with the gravity on the three dimensional AdS space as 
the simplest example where the AdS/CFT correspondence can be 
understood as the Ward identity.
Then we discuss its generalization to the cases of higher dimensional ones
and supersymmetric theories.
The matter fields coupled to AdS background can be discussed in the
context of dimensional reduction of the higher dimensional 
AdS Chern-Simons gravity to a low dimensional AdS space.
Since the near horizon geometry of the D-brane configurations for the
five dimensional black hole \cite{five}
is known to contain $AdS_3$ \cite{mal97,hyun,sfet},
the present work may help us to improve our understanding on the AdS duality
of these D-brane configurations \cite{malstr,martinec}. 


The three dimensional gravity in the presence of the cosmological 
constant $-1/l^2$ is described by a Chern-Simons theory 
with a gauge group $SL(2,R) \otimes SL(2,R)$ \cite{witten88}.
\begin{eqnarray}
I_{CS}(A,\bar A) &=& \frac{k}{4\pi} \int_M \, {\rm tr} \, \left(AdA +
\frac{2}{3}AAA \right) - \frac{k}{4\pi} \int_M \, 
{\rm tr} \, \left(\bar{A}d\bar{A} + 
\frac{2}{3}\bar{A}\bar{A}\bar{A} \right) \label{csg}
\end{eqnarray}
where $k= -\frac{l}{4G},  \quad A = A^{a}J_a, \quad 
\bar A = \bar A^{a}\bar J_a$ and
\begin{eqnarray}
\left[ J_a , J_b \right] &=& {\epsilon_{ab}}^c J_c, \quad
\left[ \bar J_a , \bar J_b\right] = {\epsilon_{ab}}^c \bar J_c, \quad
\left[ J_a , \bar J_b \right] = 0. \nonumber
\end{eqnarray} 
In terms of the dreibein and the spin connection the gauge fields
are written as
\begin{equation}
A_\mu{}^a = \omega_\mu{}^a + \frac{e_\mu{}^a}{l}, \quad
{\bar A}_\mu{}^a = \omega_\mu{}^a - \frac{e_\mu{}^a}{l}.
\end{equation} 

The following discussions will apply to all classical
solutions which are asymptotically $AdS_3$, including the
BTZ black hole \cite{btz}.
The equations of motion for CS gravity action contain the
torsion free condition and on shell the CS gravity action coincides
with Einstein action in the presence of the cosmological 
constant.
In the presence of boundary we should supplement some boundary terms
so that the variation of the gauge fields consistently yields 
the equation of motion. In ref.\cite{tlee97} the following 
boundary term has been introduced
\begin{eqnarray}
I_B = \frac{k}{4\pi} \int_{\partial M}\, 
{\rm tr} (A_\tau - A_\phi) A_\phi-  \frac{k}{4\pi} \int_{\partial M}\, 
{\rm tr} (\bar A_\tau + \bar A_\phi) \bar A_\phi, \label{bdaction}
\end{eqnarray}
where $A_{R/L} = A_\tau \pm A_\phi$, and 
$\bar A_{R/L} = \bar A_\tau \pm \bar A_\phi$.  
This boundary term is uniquely chosen in that it
does not impose a unitarity problem.

The Chern-Simons theory in three dimensions is topological; 
if the space-time manifold has no boundary, 
the Chern-Simons theory is gauge invariant and
only has finite degrees of freedom.
But in the presence of the boundary the gauge invariance is broken and
gauge degrees of freedom corresponding to the broken symmetry
becomes dynamical \cite{carlip}.
The most efficient procedure to obtain the action for these
"would be" gauge degrees of freedom is to construct a one-cocycle
\cite{tlee97,tlee90}
\begin{eqnarray}
\alpha_{G}[A,\bar A,g,\bar g] 
 &=& I_G(A^g,\bar A^{\bar g})+I_B(A^g,\bar A^{\bar g}) 
 - I_G(A,\bar A)-I_B(A,\bar A)\\
 A^g &=& g^{-1} d g + g^{-1} A g \nonumber \\
 \bar A^{\bar g} &=& {\bar g}^{-1} d {\bar g} + 
 {\bar g}^{-1} {\bar A} {\bar g}. \nonumber
\end{eqnarray} 
The explicit expression for the one-cocycle is given as
\begin{eqnarray}
\alpha_{G}(A, \bar{A},g,{\bar g})
&=& \alpha_1(A,g)+ {\bar \alpha}_1(\bar{A},{\bar g}), \label{acsg} \\
\alpha_1(A,g) &=& -\Gamma^L[g] - 
\frac{k}{2\pi}\int_{\partial M} {\rm tr}
(\partial_\phi g g^{-1})A_{L}, \nonumber\\
\bar \alpha_1(\bar A,\bar g) &=&  \Gamma^R[\bar g^{-1}] - 
\frac{k}{2\pi}\int_{\partial M}{\rm tr} (\partial_\phi
{\bar g}{\bar g}^{-1}){\bar A}_{R}, \nonumber \\
\Gamma^{L/R}[g] &=& \frac{k}{4\pi} \int_{\partial M} {\rm tr}
(g^{-1}\partial_\mp g)(g^{-1} \partial_\phi g) +\frac{k}{12\pi}
\int_{M} {\rm tr} (g^{-1} dg)^3\nonumber
\end{eqnarray}
where $\partial_{\pm} = \frac{1}{\sqrt 2}
(\partial_\tau \pm \partial_\phi)$.
Thus, the action for the local degrees of freedom is
given by a direct sum of two chiral WZW actions
coupled to the gravity (gauge field) on the boundary. 
The complete action consists of
the Chern-Simons terms, the boundary terms and the one-cocycle
\begin{eqnarray}
I_{G} (A, \bar{A},g,{\bar g}) = I_{CS} (A, \bar{A}) +
I_B (A, \bar{A}) + \alpha_{G}(A, \bar{A},g,{\bar g}).
\end{eqnarray}
Thanks to the one-cocycle condition
\begin{eqnarray}
\delta \alpha_{G} = \alpha_{G}(A^h, \bar A^{\bar h}, g,\bar g) - 
\alpha_{G}(A,\bar A, hg, {\bar h}{\bar g}) 
+\alpha_{G}(A, \bar A, h,\bar h)=0 \label{one}
\end{eqnarray} 
satisfied by $\alpha_{G}[A,\bar A,g,\bar g]$, 
the gauge symmetry is fully restored
\begin{eqnarray}
Z &=& \int D[A,{\bar A}] D[g, {\bar g}] \exp \left[iI_G 
(A^h, \bar{A}^{\bar h},g,{\bar g})\right] \nonumber \\
&=& \int D[A,{\bar A}] D[g, {\bar g}] \exp \left[iI_G 
(A, \bar{A},hg,{\bar h}{\bar g}) \right] \label{inv} \\
&=& \int D[A,{\bar A}] D[g, {\bar g}] \exp \left[iI_G 
(A,{\bar A}, g,{\bar g}) \right]. \nonumber
\end{eqnarray}


As we have seen that the coupling of the boundary degrees of freedom
to the bulk degrees of freedom is dictated by the gauge invariance.  
This coupling is the important key element to understand the entropy
of the BTZ black hole and the central extension of the gauge 
algebra \cite{bana98,behrndt98,tlee98}. 
As we will see that the coupling term
is also important to understand the dual relationship
between the gravity in the bulk and the conformal field theory
on the boundary. With the coupling term 
the gauge symmetry $SL(2,R) \otimes SL(2,R)$ is faithfully 
respected by the system. The gauge symmetry in quantum field
theory, often expressed as the Ward identity, 
is important in many respects. Here we will find that 
the gauge symmetry is even more powerful in the 
topological field theory.

We may define the generating functional as follows
\begin{eqnarray}
Z[A|, {\bar A}|] = \int D[A] D[{\bar A}] D[g]D{\bar g}]
\exp i \left[I_{CS} (A, \bar{A}) +
I_B (A, \bar{A}) + \alpha_{G}(A, \bar{A},g,{\bar g})\right].
\end{eqnarray}
In the path integral the gauge fields are subject to the boundary
condition; $A|_{\partial M} = A|(t,\phi)$, 
${\bar A}|_{\partial M} = {\bar A}|(t,\phi)$.
Hereafter we denote collectively the gauge fields by $A$ and the conformal
fields by $g$ for the sake of convenience.
In the large $k$ limit, the classical trajectory will give
dominant contribution to the path integral. So we may approximate 
the generating functional as
\begin{eqnarray}
Z[A|] = \int D[g] \exp i \left\{I_{CS-class} (A|)
+ \alpha_{G}(A|,g)\right\} \label{approx}
\end{eqnarray}
where $I_{CS-class} (A|)$ denotes the action
evaluated at the classical solution of which boundary value
is given by $A|(t,\phi)$. It may be subject to radiative
corrections.
The classical equations for $A$ read as
\begin{eqnarray}
F_{\tau\phi} &=& 0 ,\nonumber\\
F_{\rho\phi} - A_\phi \delta\left({\rho-\rho_\infty}\right) &=& 0, 
\label{class}\\
F_{\tau\rho} + A_\phi \delta\left({\rho-\rho_\infty}\right) &=& 0 
\nonumber
\end{eqnarray}
where the surface of $\rho=\rho_\infty$ is the boundary $\partial M$. 
The general solution to the classical equations, Eq.(\ref{class}) which
can be found in ref.\cite{deser84} would be useful to evaluate
$I_{CS-class}(A|)$ in Eq.(\ref{approx}). 
Now let us take an (infinitesimal)
arbitrary variation of the boundary value $\delta A|(t,\phi)$.
Then one can find a gauge transformation, with which 
the small variation on the boundary can be compensated.
(The gauge function $u$ is only required to satisfy,
$\delta A| = idu + i [A|,u]$ on $\partial M$.)
Thus, the variation of the
boundary values of the fields is equivalent to a gauge variation.
Since the total action we construct fully respects the
gauge symmetry Eq.(\ref{inv}), it implies that the 
generating functional is independent of this variation.
This observation leads us to an important consequence.

Using the gauge invariance of the generating
functional Eq.(\ref{inv}), we find
\begin{equation}
Z[A|+\delta A|] =Z[A|].
\end{equation}
Equivalently the generating functional can be said to satisfy 
\begin{eqnarray}
\left(\frac{\delta}{\delta A|}\right)^n \ln Z[A|]=0. 
\end{eqnarray}
This identity may be rewritten as
\begin{eqnarray}
& & \frac{\delta}{\delta A|(y_1)} \cdots \frac{\delta}{\delta A|(y_n)}
I_{CS-class} (A|) \nonumber \\
&=& c \frac{\delta}{\delta A|(y_1)} \cdots \frac{\delta}{\delta A|(y_n)}
\ln \left\{\int D[g] \exp \left[i\int_{\partial M}(L(g)+ A|J)\right]\right\}
\label{iden} \\
&=& c \left< J(y_1) \cdots J(y_n) \right>_{connected}\nonumber
\end{eqnarray}
where $c$ is a constant
The RHS of Eq.(\ref{iden}) is precisely the {\it connected} Greens
function. Here $A| = A|_L$, (or ${\bar A}|_R$) and 
$J =- \frac{k}{2\pi} \partial_\phi g g^{-1}$, 
(or ${\bar J} = -\frac{k}{2\pi} \partial_\phi{\bar g} {\bar g}^{-1}$).
This is the AdS/CFT duality relationship in the gravity sector
presented in the simplest form.
For two-point correlation function we obtain
\begin{equation}
\frac{\delta}{\delta A|(y_1)}\frac{\delta}{\delta A|(y_2)} 
I_{CS-class} (A|) = c \left< J(y_1) J(y_2) \right>.
\end{equation}


In ref.\cite{witten981} scalar fields and $U(1)$ gauge fields
in the AdS background were taken as examples to illustrate the
AdS/CFT correspondence. In order to see this AdS/CFT duality in the
matter field sector we need to extend the theory.
We will take the Chern-Simons gravity in higher dimensions
for this purpose. The Chern-Simons gravity in higher dimensions
has been discussed extensively in refs.\cite{cham90,bana96,tron97}.
In $(2n+1)$ dimensions the anti-de Sitter group is $SO(2,2n)$.
Let $J_{AB}$ be the generators of the AdS group, where
$A$, $B = 0, 1,\dots,2n+1$ and $A= A^{AB}J_{AB}$ be the Lie
algebra valued one-form.
The vielbein and spin connection are identified by
\begin{equation} 
A^{a,2n+1} = e^a/l,\quad A^{ab}=\omega^{ab},\quad a=0,1,\dots,2n.
\end{equation}
Similarly, the two form $F^{AB}$ is given in geometrical terms by
\begin{eqnarray}
F^{a,2n+1} &=& T^a/l = \frac{1}{l}\left(de^a+\omega^a{}_b e^b\right) \\
F^{ab} &=& R^{ab} +\frac{1}{l^2} e^a e^b.
\end{eqnarray} 

The Chern-Simons action is constructed to be
\begin{eqnarray}
I_{CS} = k\int_{M} \omega_{2n+1} =
k \int_{M} (n+1) \int^1_0 d\tau \, {\rm tr}\, 
A\,(\tau dA+ \tau^2 A^2)^n
\end{eqnarray}
and the equations of motion for the Chern-Simons action read as
\begin{equation}
\epsilon_{A_1 B_1 \dots A_{n+1} B_{n+1}} 
F^{A_2 B_2} \dots F^{A_{n+1} B_{n+1}} = 0. \label{eqm}
\end{equation}
As in the case of (2+1) dimensional one, we need to introduce
appropriate boundary terms and boundary conditions in order to
consistently define the Chern-Simons theory. Here we will
not attempt to construct the explicit expressions for
them, which depend on the topological structure of $M$ and the
boundary conditions to be imposed.
We only point out that the boundary action can be constructed
as before, using the transformation of the Chern-Simons form under
the gauge transformation
\begin{equation}
\omega^g_{2n+1} = \omega_{2n+1}+ d\alpha_{2n}(A,g) +
c_n {\rm tr} \left(g^{-1} dg \right)^{2n+1}
\end{equation}
where $d\alpha_{2n}(A,g)$ is a $2n$ form and $c_n$ is a 
constant. By construction the boundary action will satisfies
the one-cocycle condition Eq.(\ref{one}).
Since the Chern-Simons gravity in higher dimensions
is equivalent to the Einstein gravity only in the semi-classical
regime where it is expanded around a classical solution
in the broken phase, we only consider the theory in
the large $k$ limit.
The procedure to get the topological Ward identity for the
higher dimensional Chern-Simons gravity is exactly 
the same as that for the (2+1) dimensional case.

In order to obtain the dual correspondence for the scalar and
$U(1)$ vector fields, we will take the dimensional reduction
of the (4+1) dimensional Chern-Simons gravity on
$AdS_4 \times S^1$.
the gravitational sector of the Chern-Simons gravity in (4+1) 
dimensions, expanded around the classical solution,
$AdS_4 \times S^1$, coincides that of the standard Einstein
gravity in (3+1) dimensions in the presence of the cosmological
constant \cite{cham90}.
Upon dimensional reduction, some components of the
Chern-Simons gauge field become a scalar field, ($A^{a4}{}_4$) 
and a vector field, ($A^{a4}{}_\mu$), $\mu=0,1,2,3$ in the $AdS_4$ space.
The equations of motions for these matter fields
are equivalent to those for the scalar and vector in the
$AdS_4$ background. (Since we mainly concern the theory around
the classical background where the Chern-Simons gravity is 
reduced to the Einstein gravity, we may adopt the usual 
Kaluza-Klein reduction.) For those components, corresponding to the
scalar and vector, Eq.(\ref{iden}) may reduce to the
desired dual relationship. In particular, for the scalar
$\varphi$,
\begin{equation}
\frac{\delta}{\delta \varphi|(y_1)} \frac{\delta}{\delta \varphi|(y_2)}
I_{class}(A_0;\varphi|) = \left<{\cal O}(y_1){\cal O}(y_2)\right>
\end{equation}
where $A_0$ denotes the $AdS_4 \times S^1$ background and ${\cal O}$, 
the operator coupled to the scalar on the boundary. Accordingly,
$I_{class}(A_0;\varphi|)$ is the classical action for the scalars
in the $AdS_4$ background evaluated for the given boundary value,
$\varphi|$. For the two-pont function, a free scalar
field action in the $AdS$ background can be taken as $I_{class}$.

One can generalize the construction of the bulk-boundary system to
the supersymmetric case by introducing the $AdS$ supergroup as
the gauge group. In (2+1) dimensions
one can introduce a supergroup $AdS(n,m)= OSp(n|2;R) \otimes
OSp(m|2;R)$ by supersymmetrizing both $SL(2,R)$ factors of 
$AdS_3$ group, $SO(2,2) \simeq SL(2,R)\otimes SL(2,R)$ \cite{achu89}.
In the supersymmetric theory the gauge connections are written as
$A = A^aJ_a+ a^{ij}_+ T^{ij}_+ + {\bar \psi_+}^i Q^i_+$, 
${\bar A}= {\bar A}^a{\bar J}_a+ a^{ij}_-T^{ij}_- + {\bar \psi_-}^i Q^i_-$
where $T^{ij}_+$ and $T^{ij}_-$ are the generators of the group 
$O(n)$ and $O(m)$ respectively and $Q^i_\pm$ are the supercharges. 
It would be straightforward to obtain
the boundary superconformal action by making use of the one-cocycle
construction. We note that the bosonic part of the action now
contains the $O(n)$ and $O(m)$ Chern-Simons terms in the bulk
and the $O(n)$ and $O(m)$ chiral gauged WZW terms on the boundary.
It would be interesting to explore the details of the AdS/CFT
duality of this supersymmetric model. It would serve as
a useful tool to evaluate the correlation functions of 
various operators in the $O(n)$ WZW model.
More interesting case
may be the (4+1) dimensional Chern-Simons supergravity 
\cite{cham90}, where one can introduce $SU(n)$ Chern-Simons gauge
fields. In this case we get the gauged chiral Wess-Zumino 
effective action in (3+1) dimensions as a part of the bosonic action 
on the boundary. Since the $SU(n)$ Wess-Zumino action is 
known to describe the QCD in the large $N$ limit \cite{witten83}, 
the duality of this theory may be useful to study the QCD 
in the large $N$ limit.

In the present paper we studied the (2+1) dimensional gravity
in the framework of the Chern-Simons theory. Constructing the
boundary theory as a one-cocycle, we showed that the Ward identity
leads us to the AdS/CFT duality in the simplest form.
Then we discuss its extensions to the higher dimensional 
Chern-Simons gravity and point out that AdS/CFT duality for
the scalar and $U(1)$ vectors in the AdS background may be
described in the dimensional reduction of the Chern-Simons
gravity in the higher dimensions. Although this work may be
the first attempt to derive the AdS/CFT dual relationship
from the first principle, it is neither rigorous nor 
complete. This work only suggests that the AdS/CFT duality
may be interpreted as a topological Ward identity in some simple
examples. In order to understand the $AdS_5 \times 
S^5$ duality, which is the most interesting one in the string 
theory, we may need a better framework where the dynamics 
of the string-D-brane can be properly described. 
Since the holographic field theory for M-theory recently 
proposed by Horava \cite{hora} is also based on the 
eleven dimensional AdS Chern-Simons supergravity, the present
work may be useful to explore some aspects of M-theory along this 
line. It seems that the dual relationship between 
the bulk-boundary theory is not limited to the AdS/CFT 
duality of string theory. 


This work was supported in part by the Basic Science Research Institute 
Program, Ministry of Education of Korea (BSRI-97-2401)
and by KOSEF through CTP at SNU and through Project 95-0702-04-01-3. 
I would like to thank Professor R. Jackiw for informing me of 
ref.\cite{deser84} and S. Hyun and J. H. Cho for useful discussions.

\end{document}